\documentclass[]{ceurart}

\usepackage{listings}
\usepackage{subcaption}
\begin{document}

\copyrightyear{2024}
\copyrightclause{Copyright for this paper by its authors.
  Use permitted under Creative Commons License Attribution 4.0
  International (CC BY 4.0).}

\conference{SEMANTiCS 2024 EU: 20th International Conference on Semantic Systems, September 17-19, 2024, Amsterdam, The Netherlands}

\title{ORKG ASK: a Neuro-symbolic Scholarly Search and Exploration System}

\author[1]{Allard Oelen}[%
orcid=0000-0001-9924-9153,
email=allard.oelen@tib.eu,
]
\address[1]{TIB – Leibniz Information Centre for Science and Technology, Hannover, Germany}

\author[2]{Mohamad Yaser Jaradeh}[%
orcid=0000-0001-8777-2780,
email=jaradeh@l3s.de,
]
\address[2]{L3S Research Center, Leibniz University of Hannover, Hannover, Germany}

\author[1,2]{S\"oren Auer}[%
orcid=0000-0002-0698-2864,
email=auer@tib.eu,
]

\begin{abstract}
Purpose:
Finding scholarly articles is a time-consuming and cumbersome activity, yet crucial for conducting science. Due to the growing number of scholarly articles, new scholarly search systems are needed to effectively assist researchers in finding relevant literature. 

\noindent Methodology:
We take a neuro-symbolic approach to scholarly search and exploration by leveraging state-of-the-art components, including semantic search, Large Language Models (LLMs), and Knowledge Graphs (KGs). The semantic search component composes a set of relevant articles. From this set of articles, information is extracted and presented to the user. 

\noindent Findings:
The presented system, called ORKG ASK (Assistant for Scientific Knowledge), provides a production-ready search and exploration system. Our preliminary evaluation indicates that our proposed approach is indeed suitable for the task of scholarly information retrieval. 

\noindent Value:
With ORKG ASK, we present a next-generation scholarly search and exploration system and make it available online. Additionally, the system components are open source with a permissive license. 
\end{abstract}


\maketitle

\section{Introduction}
Finding scholarly articles and exploring the body of scholarly literature consumes a significant share of a researcher's time. 
Due to the growing number of scholarly articles, this issue only becomes more apparent~\cite{landhuis2016scientific}. Current scholarly search systems passively assist users with their information needs by providing a list of relevant articles. If instead active assistance were provided, the users' information needs, such as a research question, would be answered for them.
We present ORKG ASK (Assistant for Scientific Knowledge), a new generation scholarly search and exploration system\footnote{Available online via \url{https://ask.orkg.org}}. ORKG ASK helps researchers find relevant literature and automatically extract knowledge from the retrieved literature, actively supporting researchers with their information needs. The approach consists of three main components: 1) Semantic Search, 2) a Large Language Model (LLM), and 3) Knowledge Graphs (KGs). First, the semantic search addresses the previously discussed challenge of retrieving articles based on their relevance to a specific information need. In ORKG ASK users can formulate their information need as a research question, which is entered as a search query.
Second, an LLM is leveraged to answer the research question by prompting with the context of the set of relevant articles. In addition to answers to the research question, a set of properties is extracted, among others, a summary, materials, methods, and results of the contributions described in the articles. 
Third, KGs are used to provide more fine-grained information extraction as well as for curating extraction results. This includes results filtering based on mentioned concepts in scholarly articles.

\begin{figure}[t]
    \centering
    \includegraphics[width=0.95\linewidth]{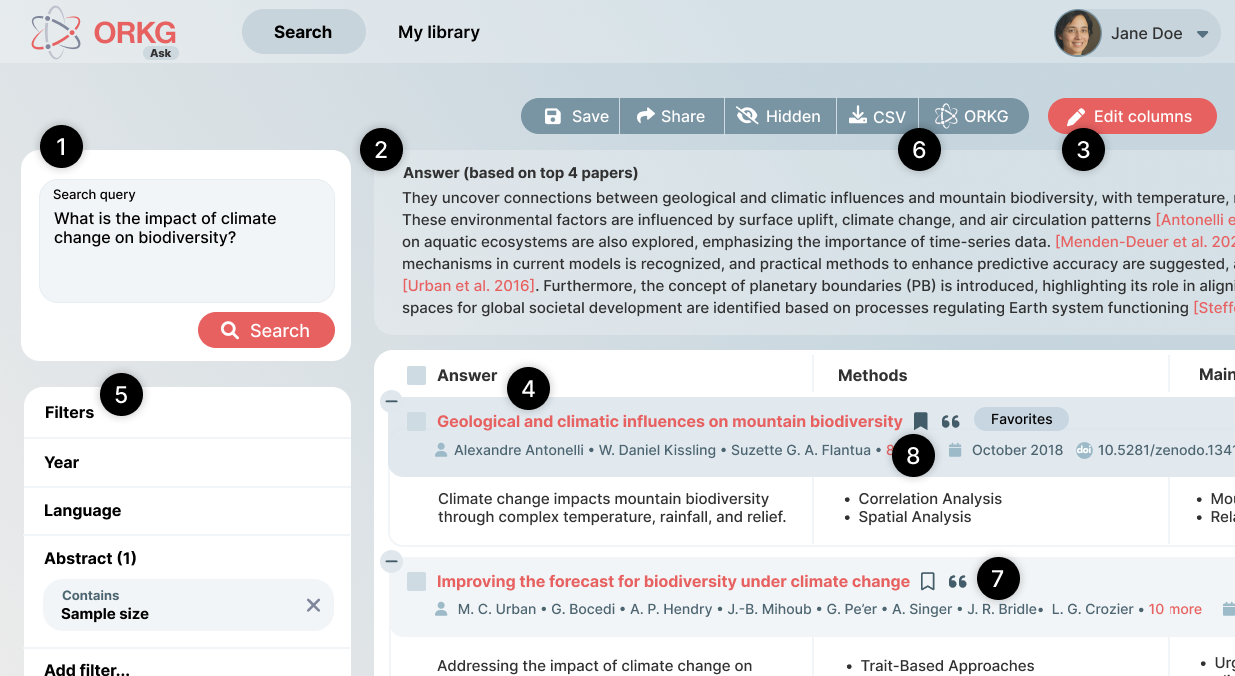} 
    \caption{Design of the search result page of the ORKG ASK application.}
    \label{fig:screenshot}
\end{figure}

\section{Background}
\label{section:related-work}
There are various established, large, and multidisciplinary scholarly search systems, among others, Google Scholar, Semantic Scholar, and Scopus~\cite{gusenbauerWhichAcademicSearch2020}.
Other systems, such as PubMed
and ACM Digital Library,
are domain-specific. These search systems take a similar approach where articles are ranked based on relevance, but where users have to manually extract relevant information from articles. A new approach provides active support via automatic information extraction by systems such as Elicit, Consensus, and Scispace~\cite{bolanosArtificialIntelligenceLiterature2024}. These systems are not open-source, leaving details about their approach, such as the model and dataset, to be unknown, in turn making results harder to reproduce. This makes such systems less suitable for systematic literature reviews where reproducibility is a key aspect of the approach~\cite{macfarlaneSearchStrategyFormulation2022}. 

To extract knowledge from a large set of scholarly documents, a Retrieval-Augmented Generation (RAG) approach can be used to provide the LLM with relevant context~\cite{lewisRetrievalaugmentedGenerationKnowledgeintensive2020}. The Retrieval aspect retrieves a set of documents, commonly done using vector databases. The Augmented aspect, augments the user query with the found context. Finally, the Generation aspect creates the response. To our knowledge, the previously mentioned AI-supported scholarly search systems use this approach and are thus similar to the approach we propose with ORKG ASK. 

\section{System Overview}

\begin{figure}[t]
    \includegraphics[width=1\linewidth]{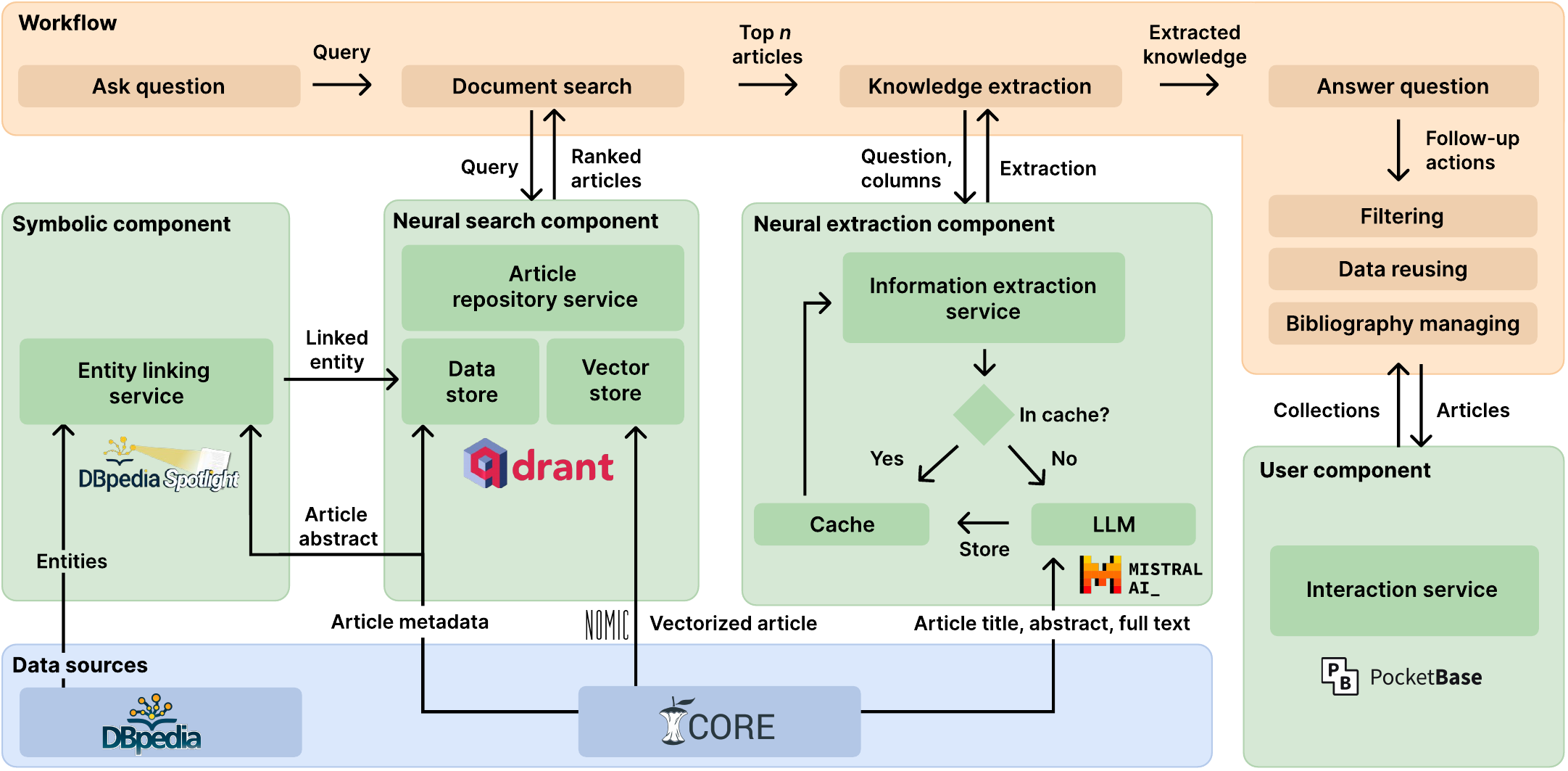} 
    \caption{ORKG ASK system workflow integrating neuro-symbolic components.}
    \label{fig:workflow}
\end{figure}

\label{section:system-overview}
We now discuss the ORKG ASK system in more detail. The service is designed and developed in such a way that it provides a solid foundation for a sustainable service. Additionally, we focus on accessibility by providing a dark mode (for low-light conditions), a responsive interface (for mobile usage, or high zoom levels for visually impaired users), and implementing ARIA accessibility attributes where necessary. The ORKG ASK code base is published as open source under an MIT license and available online.\footnote{\url{https://gitlab.com/TIBHannover/orkg/orkg-ask}} \autoref{fig:screenshot} depicts the design of the search result page for a specific research question.

\subsection{User-Oriented Features}

The \textbf{Question Answering} feature as depicted in Node 1 in~\autoref{fig:screenshot} illustrates the input field for the research question. Nodes 2 and 4 present the answers to the question. Node 2 shows a synthesized answer of the first five displayed results. The \textbf{Information Extraction} feature extracts additional information from an article (node 4). There are several default columns displayed, but users can customize the extracted information to their needs (node 3). The \textbf{Filtering} feature enables users to filter articles based on user-provided criteria (node 5). This includes the ability to filter based on year, language, words that appear in the title or abstract, the number of citations, author names, etc. The \textbf{Bibliography Managing} featured called ``My Library'' provides a bibliography manager where users can store and curate a list of articles. Articles are added by clicking on the bookmark icon in the interface (node 8) or added manually via the My Library page (via DOI, title, or BibTeX). Articles from My Library can be manually added to a search query, which prepends the manually selected articles to the search results. The \textbf{Data Reuse} feature supports citing articles in APA, Vancouver, Harvard, citation styles, and exports to BibTeX, RIS, and CLS-JSON (node 7). The export button is displayed in node 8. Furthermore, there is an option to export the entire search result table to CSV and ORKG~\cite{auer2020improving} CSV (node 6). Finally, the \textbf{Entity Linking} feature links entities in article abstracts to their respective DBpedia entries. This provides the ability to filter articles based on semantically identical concepts, providing an additional means to more targeted information retrieval.

\subsection{System Workflow}
\autoref{fig:workflow} depicts the system workflow. It starts with a user asking a research question. A set of relevant documents for this question is retrieved. The neural search component uses vectorized representations of the query and articles via the Nomic embeddings model to retrieve a set of relevant documents. Optionally, the search space can be narrowed down by filtering specific metadata or linked entities. Qdrant\footnote{\url{https://qdrant.tech}} is used as a vector and data store. The symbolic component processed article abstracts offline and stored these linked entities in the data store. The entity linking is conducted using DBpedia Spotlight~\cite{mendes2011dbpedia}. The CORE dataset~\cite{Knoth2023-zi}, containing article metadata, abstracts, and full-text (in the case of open-access articles), is used as a data source for the vector store. Next, knowledge is extracted using an LLM from the top \textit{n} articles, resembling the RAG approach. Currently, we use the Mistral Instruct 7B v0.2 model for the information extraction. 
To reduce system resource usage, the LLM is only prompted if the answer does not yet exist in the cache.
Finally, the information is presented to the user. 

\begin{figure}[t]
        \centering
        \includegraphics[width=0.8\textwidth]{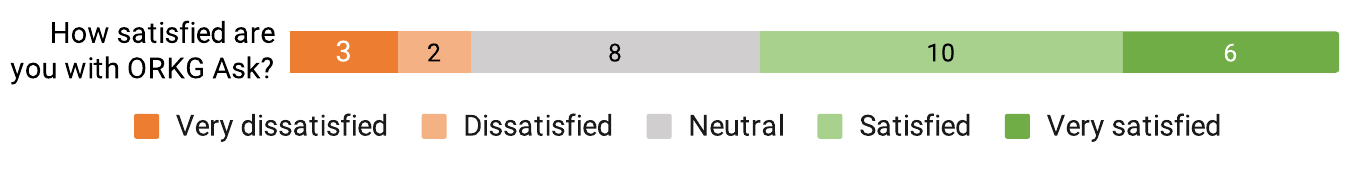}
        \caption{Results for user satisfaction evaluation indicating relatively satisfied users.}
        \label{fig:satisfaction}
\end{figure}

\section{Evaluation}
\label{section:evaluation}
\begin{figure}[t]
        \centering
        \includegraphics[width=.78\textwidth]{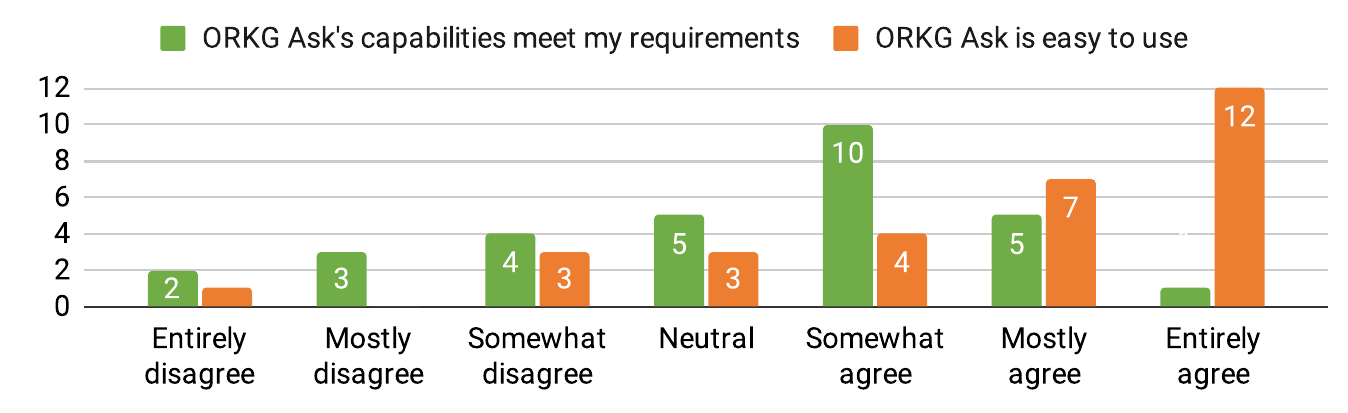}
        \caption{Results for UMUX-Lite evaluation with a total score of 65.2.}
        \label{fig:umux}
\end{figure}

As a preliminary system evaluation, we aim to assess the usability of the system. We did this using a 5-point user satisfaction assessment and the Usability Metric for User Experience lite (UMUX-lite)~\cite{lewis2013umux} evaluation. Participants were recruited via the ORKG ASK production system, via a non-intrusive tooltip asking real-world system users for their opinion. To keep participation efforts as low as possible, no participant demographics were requested from users. In total, 30 participants took part in this evaluation. As \autoref{fig:satisfaction} shows, users are relatively satisfied with ORKG ASK. The UMUX-Lite evaluation displayed in \autoref{fig:umux} results in an overall score of 65.2. As the individual results show, most participants agree that ORKG ASK is easy to use, but the system does not always meet their requirements. This could be explained by users' search behavior, as logs of asked questions revealed that not all questions are valid and answerable, leaving the user's specific search requirement unmet. Further evaluation is needed to determine what is needed to understand the user's expectations and needs better.

\section{Conclusion}
The introduction of ORKG ASK serves as a starting point for a neuro-symbolic approach to finding and exploring scholarly articles. The preliminary evaluation indicates that our approach is easy to use. In the future, we plan to extend the system by providing provenance information to highlight the source of extracted information. Furthermore, we plan to extend the KG part significantly, growing the KG automatically while the system is being used. 

\label{section:discussion-conclusion}

\bibliography{refs}

\end{document}